\documentclass[twocolumn,showpacs,preprintnumbers,amsmath,amssymb]{revtex4}

\usepackage{graphicx}
\usepackage{dcolumn}
\usepackage{bm}
\input epsf

\begin{document}

\title{Towards a statistical theory of transport by strongly-interacting lattice
fermions}
\author{Subroto Mukerjee}
\email{mukerjee@princeton.edu}
\author{Vadim Oganesyan}
\email{voganesy@princeton.edu}
\author{David Huse}
\email{huse@princeton.edu} \affiliation{Department of Physics,
Princeton University, Princeton NJ 08544}
\date{\today}

\begin{abstract}
We present a study of electric transport at high temperature in a
model of strongly interacting spinless fermions without disorder.
We use exact diagonalization to study the statistics of the energy
eigenvalues, eigenstates, and the matrix elements of the current.
These suggest that our nonrandom Hamiltonian behaves like a member
of a certain ensemble of Gaussian random matrices. We calculate
the conductivity $\sigma(\omega)$ and examine its behavior, both
in finite size samples and as extrapolated to the thermodynamic
limit.  We find that $\sigma(\omega)$ has a prominent
non-divergent singularity at $\omega=0$ reflecting a power-law
long-time tail in the current autocorrelation function that arises
from nonlinear couplings between the long-wavelength diffusive
modes of the energy and particle number.
\end{abstract}

\pacs{Valid PACS appear here}
\maketitle

\section{\label{sec:intro}Introduction}
The subject of transport in strongly-correlated electron systems
is of considerable current interest, motivated, for example, by
recent experiments on cuprate and cobaltate materials measuring
electrical and thermal transport properties \cite{Wang,HTSC}.
Traditionally, much of the work has
focused on the behavior of quantum systems at low temperatures
\cite{GroundState},
where transport properties can be used to characterize a
material's ground state. Under these conditions, the excitations
responsible for transport are often dilute and weakly interacting
(``quasiparticles"), thus allowing for a straightforward
description of transport, e.g. using the Boltzmann (kinetic)
equation formalism \cite{AGD}. Interestingly, in the opposite
extreme, when temperature is high and no well defined elementary
excitations exist, transport properties remain nontrivial and
difficult to calculate and understand \cite{highTnumerics}.
Additionally, what makes studying this regime worthwhile is the
fact that sometimes interesting phenomena take place over a broad
temperature range, extending up to temperatures so high that
direct interpretations in terms of low temperature universal
properties (either conventional or exotic) can be questioned, thus
leaving the high temperature regime addressed in this work as
 a promising starting point of analysis.
Although we attempt no detailed comparisons with experiments here,
it may be of interest to note that the optical conductivity we
find is very broad, extending over frequencies of order the bare
bandwidth, with DC resistivity growing linearly with temperature.
These features are remarkably reminiscent of what is often
observed in the normal state
 of correlated materials as diverse as organic salts, CDW systems, $C_{60}$
 and even high temperature superconductors. However, further work
is necessary to elucidate possible implications of our
 high temperature approach to any specific material.

In this paper, we study the conductivity, $\sigma$, as a function
of frequency $\omega$ and temperature $T$. For a finite quantum
system with a discrete spectrum, the real part of the conductivity
at nonzero frequency is given by the Kubo formula:
\begin{equation}
\sigma(\omega,T) = \pi\frac{1-e^{-\beta \hbar \omega}}{\omega
Z}\sum_{n,m}e^{-\beta E_n}|J_{nm}|^2 \delta (E_n - E_m - \hbar
\omega)~, \label{kubo}
\end{equation}
where $n$ and $m$ are eigenstates of the Hamiltonian, $E_n$ and
$E_m$ are the corresponding energy eigenvalues, $J_{nm}$ is the
matrix element of the total current operator between these two
states, $\beta = 1/k_BT$ and $Z$ is the partition function.
Note that we do not include any ``external'' dissipative bath; what we are studying is the absorption of energy from an infinitesimal applied AC electric field by transitions between eigenstates of the Hamiltonian.

One thing we investigate here is how the above singular expression
for the conductivity converges to a continuous function in the
thermodynamic limit.  It is a sum of delta-functions, one for each
pair of energy eigenstates that are connected by the current
operator.
The number of eigenstates
scales exponentially in the number of degrees of freedom in the
system  $\sim e^{sL^d}$, where $L$ is system's linear extent and
$s$ is the entropy density, with average level spacing $\Delta\sim
e^{-sL^d}$. For a generic non-integrable Hamiltonian there are
only a few good quantum numbers.  These include the total number
of particles, the Bloch (lattice) momentum for systems with
discrete translational invariance, and in some models of interest
the total spin.  All pairs of states with the same quantum numbers
generically contribute a delta-function to the conductivity, and
so the total number of terms in the sum scales as $\sim
e^{2sL^d}$. Thus the number of delta-functions in
$\sigma(\omega)$ grows very rapidly both with increasing system
size and with increasing temperature (and thus $s$), and away from
zero frequency a good approximation to the thermodynamic limit is
rapidly approached with only very fine binning or other smoothing
of the delta-functions.  However, there are significant and
interesting finite size effects present near zero frequency, and
elucidating them is one of the main topics of this paper.

For models where a finite-sized system has a bounded spectrum and
a finite-dimensional Hilbert space, such as the one-band model of
interacting lattice fermions that we will study here as our
example Hamiltonian, the above expression simplifies in the high
temperature limit to
\begin{equation}
k_BT\sigma(\omega) \approx \frac{\pi\hbar}{Z}\sum_{n,m}|J_{nm}|^2
\delta (E_n - E_m - \hbar \omega)~. \label{hikubo}
\end{equation}
We will mostly focus on this limiting behavior of the
conductivity, since it is already nontrivial.  However, all of the
features we will discuss appear to apply at finite temperatures as
well.  One basic noteworthy feature of this high $T$ limit is that
the resistivity is proportional to the temperature.

Such a simple statistical characterization of the conductivity
breaks down at low temperatures and frequencies.  At low
temperatures both thermodynamics and dynamics are dominated by
comparatively few transitions from low energy states residing in
the tail of the Hamiltonian's full spectrum, and their properties
may depend on specific details of the system and its ground state.
The low frequency part of the high temperature conductivity, on
the other hand, shows various finite size effects that can be
understood based on rather general considerations:

i) Level repulsion reduces the number of pairs of states with a
very small energy difference, producing an underestimate of the
macroscopic low-frequency conductivity (in fact, $\sigma$ vanishes
as $\omega\rightarrow 0$);

ii) If the above formula (1) for the conductivity of a finite
system is applied precisely {\it at} $\omega=0$, we find a delta function
due to the current's diagonal matrix elements $J_{nn}$.
For integrable Hamiltonians this delta-function has a nonzero
weight in the thermodynamic limit, hence the macroscopic DC
conductivity appears to be infinite at all temperatures.  In
weakly non-integrable systems this delta function is weakly
broadened, it is the celebrated Drude peak.  In our case, and in
non-integrable systems more generally, the weight of this
precisely elastic delta-function in the conductivity scales
exponentially to zero with increasing system size, as discussed in
Section IV below.

iii) Finally, in the infinite system there is a non-divergent
zero-frequency singularity of the form
$\sigma(\omega=0)-\sigma(\omega)\sim \omega^{d/2}$ due to
nonlinear couplings between the long-wavelength diffusive energy
and particle density modes.  This singularity is rounded by
finite-size effects.

We discuss and disentangle all of these effects (as well as
demonstrate the general validity of considerations away from
$\omega=0$) using a simple one dimensional model as a case study.
While some of the results are special to the particular model,
e.g. the dependence of the low frequency non-analyticity on
dimension and range of interactions, the overall picture that
emerges is clear and robust against variations in parameters, and
most features are not special to one dimension.  It also appears
to apply at all temperatures in the disordered phase provided
$\beta \omega \ll1$.  In Section II we present the Hamiltonian
under study as well as the numerical procedures employed to solve
the problem exactly for finite sizes. In Section III we use the
exact numerical results to study statistical properties of the
spectrum and the eigenstates, including the matrix elements of the
current operator. The conductivity is computed and discussed in
Section IV, and its diffusion-induced non-analyticity is analyzed
in Section V.  Finally, we conclude with a brief overview and list
of open questions.

\section{\label{sec:diagonalization}$t-t'-V$ model and exact diagonalization}
Although much of what we find is rather generic to non-integrable
quantum many-body solid-state systems, the specific model we
consider here is a one dimensional chain of spinless fermions with
a single tight-binding orbital at each site.  The Hamiltonian is
\begin{equation}
H=-t\sum_{j}c_j^+c_{j+1} - t'\sum_{j}c_j^+c_{j+2} + {\rm h.c.} +
V\sum_{j}n_{j}n_{j+1}~, \label{ttV}
\end{equation}
where $c_j^+$, $c_{j}$ and $n_j=c_j^+c_{j}$ are fermion creation,
annihilation and density operators, respectively. This is one of
the simplest non-integrable (quantum chaotic) many-particle
models.  For spinless lattice fermions the size of the Hilbert
space is small, with only 2 states per site, and thus $2^L$ states
for a chain of length $L$ sites.  We have a short-range
(nearest-neighbor) interaction $V$ (spinless lattice fermions
cannot have an on-site interaction). The second-neighbor hopping
$t'$ makes the Hamiltonian non-integrable and also breaks
particle-hole symmetry. This system's good quantum numbers
(conserved quantities in addition to the energy) are total
particle number and total crystal momentum. As written above, with
the periodic boundary conditions that we use, it also has parity
symmetry. We have chosen to break parity with an irrational phase
twist at the boundary (equivalent to threading a fraction of a
magnetic flux quantum through the loop). This minimizes the number
of distinct symmetry sectors to those labeled by the two good
quantum numbers. We have also looked at the model where the
second-neighbor hopping is replaced by a second-neighbor
interaction; this latter model is equivalent to that studied by
Rabson, {\em et al.} \cite{rabson}. We chose to focus instead on
our model because we did not want particle-hole symmetry. However,
the results we present are generally independent of the presence
or absence of any of these symmetries other than conserved
particle number. Similar models have been used in earlier studies
of charge, heat and/or spin transport and the crossover from
integrability to non-integrability \cite{rabson, highTnumerics}.

For this model the operator for the total particle current is
\begin{equation}
J=\frac{it}{\hbar}\sum_{j}c_j^+c_{j+1} + \frac{2it'}{\hbar}
t'\sum_{j}c_j^+c_{j+2} + {\rm h.c.} \label{ttV}
\end{equation}
We have also looked at heat transport and thermopower, with
similar preliminary results; this work will be reported later.  We
work for convenience at half filling, and average over the
different total crystal momentum sectors.  It appears, as we
expect, that the behavior is statistically the same in each such
momentum sector, and that there is nothing special about half
filling for the regimes and properties we examine. We specifically
take $t=t'=1$ and $V=2$, so the interaction is roughly equal to
the single-particle bandwidth, and thus strong. All energies are
measured in units of $t=t'=1$, frequencies in units of
$t/\hbar=1$, and lengths in units of the lattice spacing.

We use a Householder diagonalization routine to exactly
diagonalize the Hamiltonian within each momentum sector (typically
16 or 18 sites, with 8 or 9 particles, respectively).  For
convenience, we choose to work in the basis of single-particle
momentum eigenstates.  These states are also eigenstates of the
total current operator, so this is one of the two ``natural''
bases for this problem.  The other natural basis is less simple;
it is that of the eigenstates of the other operator that enters in
determining the conductivity, namely the Hamiltonian itself.

\section{\label{sec:random}Statistics of the spectrum, eigenstates and current matrix elements}
One of the common measures used to characterize the spectrum of a
non-integrable quantum Hamiltonian is the (Wigner-Dyson)
distribution of level spacings. Our Hamiltonian is a real matrix
(this is true in our basis of single-particle momentum eigenstates
even with a magnetic flux through the loop), so the probability
distribution of the level spacing should be that of the GOE
ensemble of random matrices,
\begin{equation}
P(s) = {{\pi s}\over{2}}e^{-\pi s^2/4}~, \label{level}
\end{equation}
where $s$ here is the level spacing in units of the mean level
spacing at that energy.  To compare to this prediction we first
have to measure the density of states, which sets the mean level
spacing. The average density of states $N(E)$ in each sector of
total crystal momentum is shown in Fig. 1 for the case of chain
length $L=18$.  Thus we obtain an energy-dependent mean level
spacing $\Delta(E)=1/N(E)$.  The spacing of each pair of adjacent
levels within one sector is divided by $\Delta(E)$ to yield the
scaled level spacing $s$. The resulting probability distribution
$P(s)$ for $L=18$ is shown in Fig. 1, where we can see it agrees
very well with the GOE form (\ref{level}).
\begin{figure}[h!]
\begin{center}
$\begin{array}{c} \label{levelsp} \epsfxsize=3in
\epsffile{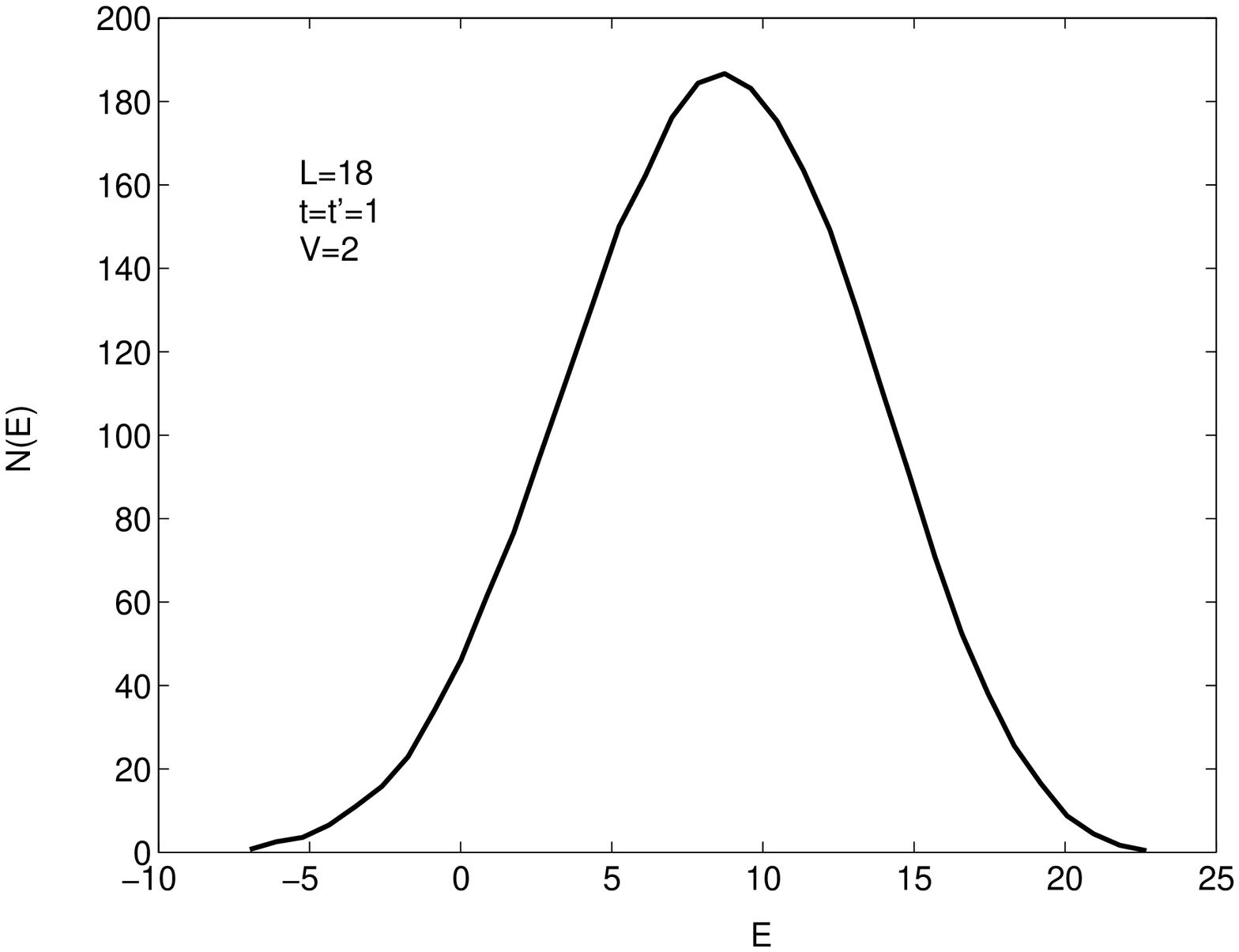} \\
\epsfxsize=3in
\epsffile{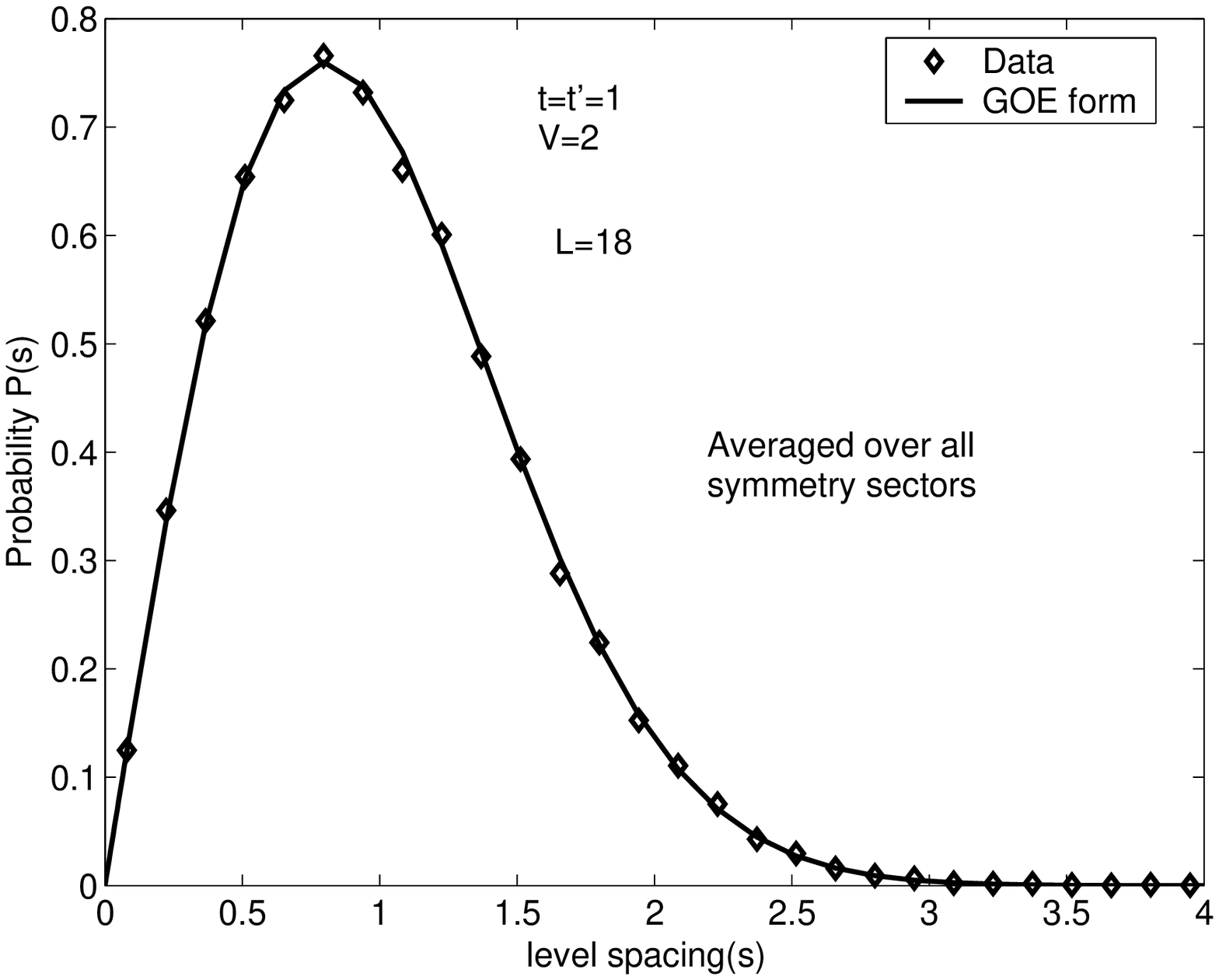} \\
\end{array}$
\end{center}
\caption{(Top) The average density of states in each symmetry
sector.  (Bottom)  The probability distribution of the scaled
level spacing for the $L=18$ system (data points) and the GOE form
for this distribution (line), showing the excellent agreement.}
\end{figure}

The Kubo formula (\ref{kubo}) shows that a relevant statistical
quantity that enters the conductivity at a given frequency
$\omega$ is the mean-square value of the current matrix elements
$J_{mn}$ between energy eigenstates separated by an energy
$\hbar\omega=|E_m-E_n|$.  [Note that our eigenstates all have real
amplitudes, so these matrix elements are all real here.]
Thus we were interested in characterizing the probability
distribution of these current matrix elements.  What we find is
that the distribution is consistent with a Gaussian, with mean
zero and a variance that depends on both the initial and final
energies. The variance has a strong dependence on $\omega$, giving
$\sigma(\omega)$ its dependence on $\omega$ (see below).  The
variance also has a weaker dependence on the average energy $\bar
E=(E_m+E_n)/2$.  To examine the probability distribution of
$J_{mn}$, we bin the matrix elements by both $\omega$ and $\bar
E$, and look at all the $J_{mn}$'s that fall in one such bin of
energies, accumulated over all total momentum sectors. We measure
the first through fourth moments, $M_1$ through $M_4$, of these
distributions, observing that the odd moments are consistent with
zero, and looking particularly at the ratio $R=M_4/M_{2}^{2}$ that
is equal to 3 for the Gaussian distribution.  If the bins are
chosen too wide, the variance changes over the bin and the
resulting distribution has a ratio $R$ larger than 3.  But as the
bins are made narrower, $R$ converges to 3, indicating that the
current matrix elements do indeed have Gaussian probability
distributions.  Since the variance depends much more strongly on
$\omega$ than on $\bar E$, the bins are taken to be much narrower
along the $\omega$ direction.

Thus we conclude that the total current operator, as written in
the basis of the eigenstates of the Hamiltonian, is effectively a
Gaussian random matrix that is, loosely speaking, ``banded'',
meaning the variance of the matrix elements $J_{mn}$ depends
smoothly on $E_m$ and $E_n$.  The ``bands'' of constant variance
in this case do not run exactly parallel to the diagonal of the
matrix.  Of course, neither our Hamiltonian nor the current
operator contain any random variables; the apparent randomness
comes from the statistical properties of the highly excited
eigenstates of our nonintegrable (and thus ``quantum chaotic'')
Hamiltonian.  The total current operator, on the other hand, is
integrable, with its eigenstates being the simultaneous
eigenstates of all the single-particle momenta.  When we write the
Hamiltonian as a matrix in the basis of the eigenstates of the
current operator, which is precisely what we do to diagonalize it
numerically, it is a highly regular sparse matrix.  [We have also
looked at the {\it energy} current operator, which enters in the
thermal conductivity and thermopower; the energy current operator
is nonintegrable, and the Hamiltonian does look like a Gaussian
random matrix when written in terms of the eigenstates of the
total energy current.]

The final statistical characterization we have done is of the
amplitudes $\langle n|\alpha\rangle$ when the eigenstates
$|n\rangle$ of $H$ are written in the basis of the eigenstates
$|\alpha\rangle$ of $J$ (and {\it vice versa}).  These amplitudes,
like the current matrix elements $J_{mn}$ discussed above, appear
to be mean-zero Gaussian random variables with a variance
depending on two energies, namely $E_n$ and the kinetic energy
$K_{\alpha}$ of the current eigenstate.  Each current eigenstate
is also an eigenstate of the kinetic energy operator, which is the
sum of the hopping terms in the Hamiltonian.  As expected, a state
$|n\rangle$ of high (low) total energy $E_n$ consists primarily of
states $|\alpha\rangle$ of high (low) kinetic energy $K_{\alpha}$,
but it appears to be quite uniformly and randomly extended over
all the current eigenstates at each particular kinetic energy.
Thus we find that both the eigenstates of $H$, and their current
matrix elements appear to behave as Gaussian random variables with
energy-dependent variances.

\section{\label{sec:conductivity}The conductivity}
\begin{figure}[h!]
\begin{center}
$\begin{array}{c} \epsfxsize=3in
\epsffile{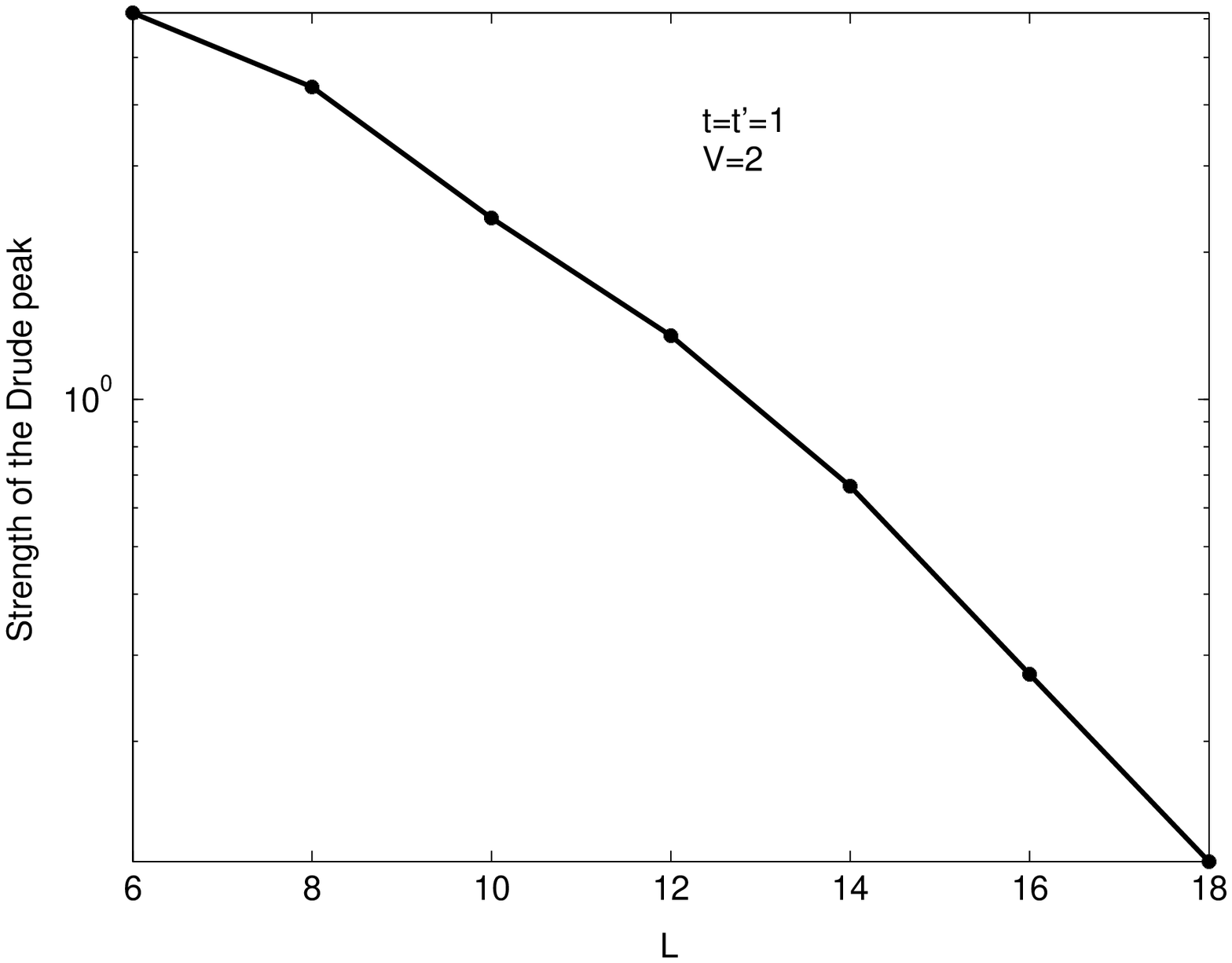}\\
\epsfxsize=3in
\epsffile{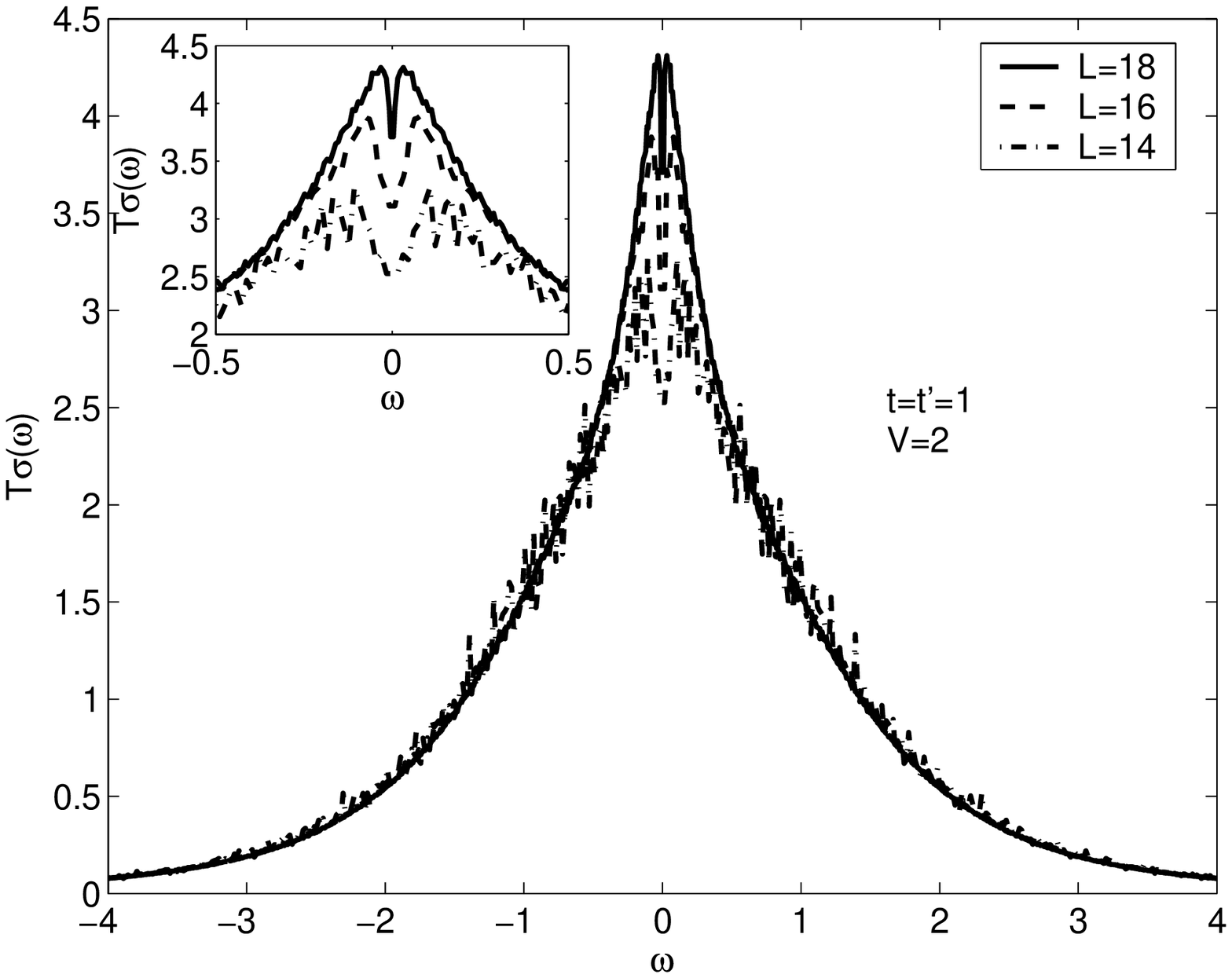} \\ 
\end{array}$
\end{center}
\caption{(Top) Drude weight as a function of $L$. The Drude weight
is expected to vanish exponentially with $L$. (Bottom)
Conductivity as a function of frequency for $L=18$, $L=16$ and
$L=14$.  The agreement among different system sizes is good at
high frequencies but significant finite-size effects can be seen
at low frequency. (Inset) Closer view of the low frequency regime.
The dip in the conductivity around $\omega=0$ is due to level
repulsion. This dip rapidly narrows with increasing system size.
The bins used here have a width equal to the mean level spacing at
the maximum of the density of states for that $L$, except in the
case of the main panel for $L=14$, where the bins widths are
instead chosen proportional to $\omega$ in order to reduce the
scatter.} \label{drude}
\end{figure}
The conductivity is given by the Kubo formula (\ref{kubo}). As
remarked earlier, it is a series of delta functions for a
finite-sized system, which can be binned to produce a smooth curve
for $\sigma(\omega)$.  Since the limit of zero frequency is of
particular interest, we will focus much of our attention on low
frequencies.  Although the expression (1) does not apply at
strictly zero frequency, it does contain a delta-function there
that all levels contribute to.  This zero-frequency feature has
often been called (ambiguously) the ``Drude peak'', and appears to
have finite weight in the thermodynamic limit for integrable
models.
However, for nonintegrable quantum chaotic models such as we study
here, the weight in this strictly elastic Drude peak vanishes
strongly in the thermodynamic limit, see Fig. 2, top panel.  This
can be roughly understood by noting that the full conductivity
(1), which has a finite large $L$ limit, is a sum of $\sim
e^{2sL}$ delta functions, where $s$ here is the entropy density.
Of these, only $e^{sL}$ are at zero frequency (since there are no
degeneracies within any symmetry sector).  We find that the weight
of the zero frequency terms in the sum are of the same order as
those at small frequency, so the relative strength of the
so-called Drude peak at zero frequency drops as $e^{-sL}$ with
increasing $L$.  A more careful analysis gives power-law in $L$
prefactors to these exponential dependences.  But the presentation
on a semilog plot in Fig. 2 shows that our results are consistent
with a Drude weight that vanishes exponentially with increasing
$L$.

The conductivity is shown in Fig. 2 for our longer chains, of
length 14, 16 and 18.  To obtain fairly smooth functions, the
delta functions are binned (see caption), and the elastic
$\omega=0$ ``Drude'' peak is left out of these plots.  At high
frequencies, the finite size effect is clearly very small, so
these short chains give a good approximation to the thermodynamic
limit.  An interesting question that we do not yet have an answer
for is: What is functional form of the conductivity at high
frequencies (say, $\omega>t$), and what is the physics determining
it?  We find that $\sigma(\omega)$ appears to decrease faster than
any power of $\omega$ in this regime.

While there appears to be good convergence to the thermodynamic
limit of $\sigma(\omega)$ at high frequencies for these chain
lengths, clear finite-size effects are seen at low frequencies. We
find two sources of these finite-size effects: level repulsion,
and diffusive ``long-time tails''.  The former is expected and has
been discussed before in this and many other contexts, but the
latter appears to be a new observation for this type of quantum
many-body system.  Level repulsion causes a reduction of
$\sigma(\omega)$ at frequencies of order or less than the mean
level spacing, because there are fewer pairs of levels with those
energy differences that can produce the dissipation.  The level
spacing vanishes exponentially in the system size, so this
feature, which appears as a dip in $\sigma(\omega)$ around zero
frequency, narrows very rapidly with increasing $L$, as is
apparent in the inset in Fig. 2.

\begin{figure}[h!]
\begin{center}
$\begin{array}{c} \epsfxsize=3in
\epsffile{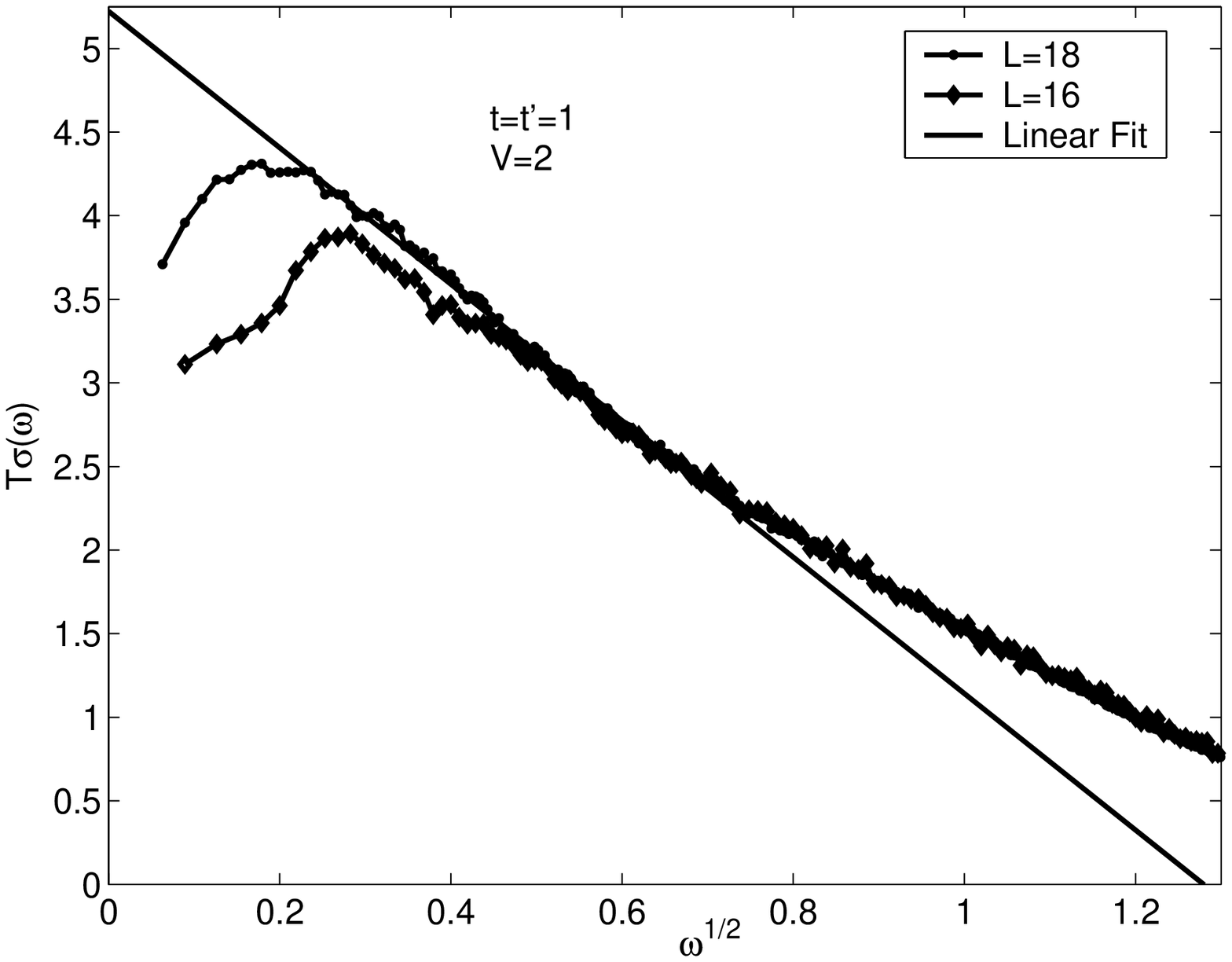}\\
\epsfxsize=3in
\epsffile{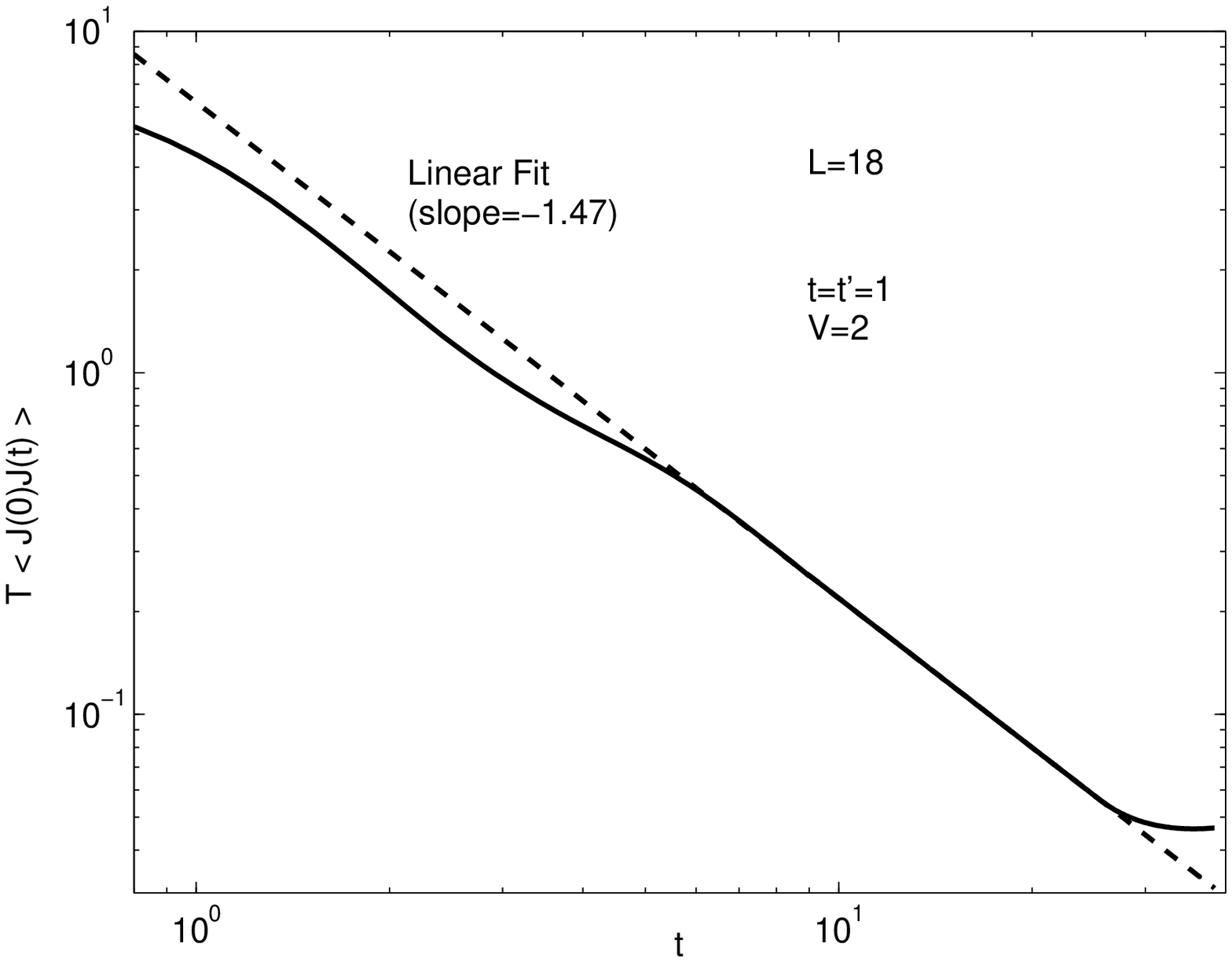} \\ 
\end{array}$
\end{center}
\caption{(Top) Conductivity plotted versus $\sqrt{\omega}$ for
$L=16$ and $L=18$. The straight line is a linear fit on this plot.
The conductivity appears linear in $\sqrt{\omega}$ at low
frequencies before it rounds off at very low frequency due to
finite-size effects. The finite-size effects are, as expected,
more pronounced for $L=16$ than $L=18$. (Bottom) Autocorrelation
function of the total current vs. time on a log-log plot. There is
a long-time tail with $\langle J(0)J(t)\rangle\sim t^{-3/2}$.}
\label{fourier}
\end{figure}

Naively, one might expect (as we did when we started this project)
that in this high temperature disordered regime with no long-range
spatial correlations the conductivity of the infinite system would
be a smooth analytic function of frequency in the vicinity of zero
frequency.  However, examination of our results in Fig. 2 show
that this does not appear to be the case.  The conductivity
appears to have a sharp maximum at $\omega=0$ that gets steadily
sharper as $L$ is increased.  This sharp peak is apparent in
published numerical data from other one-dimensional quantum models
\cite{highTnumerics}, but the reason for it appears not to have
been discussed. As we show in the next section, long-wavelength
diffusive modes of the energy and particle density generically
interact nonlinearly to produce a nondivergent zero-frequency
singularity of the form $\sigma(\omega)=a-b\sqrt{|\omega|}+...$ in
these one-dimensional systems.  In the top panel of Fig. 3 we show
that our data appear to be consistent with such a form for the
low-frequency conductivity in the large $L$ limit.  This behavior
arises from a ``long-time tail'' in the autocorrelation function
of the total current that decays with time difference as $\langle
J(0)J(t)\rangle\sim t^{-3/2}$ at long times.  We show in the
bottom of Fig. 3 that our data for this autocorrelation for $L=18$
do show such a power-law decay over a significant time range.

\section{\label{sec:hydrodynamics} Long-time tails due to diffusive modes}
Here we first give a more general argument inspired by our
one-dimensional system. Consider an infinite many-body system in
a disordered phase.  If it has any conserved densities, such as
energy, momentum, angular momentum, particle number etc., we may
ask about the transport properties for these quantities.  A
long-wavelength disturbance in the conserved quantities relaxes at
a (possibly complex) rate $\Gamma({\bf q})$ where ${\bf q}$ is the
wavevector of the disturbance, and this rate is in general
different for different linear combinations of the various
conserved quantities.  In the absence of long-range interactions
or propagating modes, the relaxation is diffusive: $\Gamma \sim
Dq^2$ for small $q$.  These long-wavelength diffusive modes in
general produce singular behavior of the transport properties in
the zero-frequency (long-time) limit, as discussed, for example,
by Kirkpatrick, {\em et al.} \cite{kirkpatrick}. This has been
discussed in the context of fluids \cite{dorfmann} where momentum
conservation makes these effects quite strong, and in systems
where quenched randomness appears to play an essential role
\cite{kirkpatrick}.

For the Hamiltonian that we are studying, there are two
conservation laws that give diffusive modes: energy and particle
number. The lattice ({\it umklapp}) breaks the momentum
conservation down to just the conservation of crystal momentum,
which does not appear to produce any slow modes in the disordered
phase.

To be a little more general, let us first consider a
translationally- and rotationally-invariant system in a disordered
phase with some number of conserved densities $n_{\alpha}({\bf r},
t)$ in possibly more than one dimension. The conservation laws
then dictate
\begin{equation}
\frac{\partial n_{\alpha}}{\partial t}=-{\bf \nabla\cdot
j}_{\alpha}~,
\end{equation}
where ${\bf j}_{\alpha}({\bf r}, t)$ is the current of conserved
quantity $\alpha$.  Let us define the local densities $n_{\alpha}$
as the deviations from the average densities, so that our system
has zero average densities $\langle n_{\alpha}\rangle = 0$. We
assume the long-wavelength, low-frequency dynamics is diffusive,
and expand the equilibrium coarse-grained dynamics in the
densities and the wavevectors (spatial gradients). The currents
are expanded as
\begin{equation}
{\bf j}_{\alpha}=-D_{\alpha\beta}{\bf \nabla}
n_{\beta}+E_{\alpha\beta\gamma}n_{\beta}{\bf \nabla}
n_{\gamma}+...+{\bf \eta}_{\alpha}~,
\end{equation}
where $D$ is the diffusivity matrix, $\eta$ is noise (due to
nonlinear couplings to short-wavelength modes) with an intensity
given by a fluctuation-dissipation relation (e.g.,
\cite{hohenberg}), and repeated subscripts are summed over.  The
lowest-order (in density and gradient) nonlinear correction to the
diffusive behavior has also been shown explicitly; for example the
term $E_{\alpha\beta\gamma}$ corresponds to the linear dependence
of the local diffusivity matrix element $D_{\alpha\gamma}$ on the
local density $n_{\beta}$.

What should be the low-frequency behavior of a transport
coefficient like $\sigma(\omega)$?  By the Kubo formula, this is
given by the time fourier transform of the autocorrelation
function of the total (zero wavevector) current. Integrating the
above expression, and thus leaving out any terms that are total
derivatives, the total current is
\begin{eqnarray}
& {\bf J}_{\alpha}(t) = \int{\bf j}_{\alpha}d{\bf r} = \nonumber \\
&
\int[\frac{E_{\alpha\beta\gamma}-E_{\alpha\gamma\beta}}{4}(n_{\beta}{\bf
\nabla} n_{\gamma}-n_{\gamma}{\bf \nabla} n_{\beta})+...+{\bf
\eta}_{\alpha}]d{\bf r}~.
\end{eqnarray}
Note that the total current is correlated with the densities only
through the nonlinear corrections to simple diffusion; we will
consider the generic case when these corrections are indeed
nonzero.  Through this nonlinearity, the zero-wavevector current
is coupled to the slow, small-wavevector diffusive density modes;
this is what produces the long time tails.  However, since the
leading nonlinear coupling that we show explicitly here must be to
an antisymmetric combination of the conserved densities, for a
system with only one conserved density (and thus no antisymmetric
combinations of densities), one has to go to significantly
higher-order nonlinearities ($|\nabla n|^2{\bf \nabla}n$) to
obtain the leading long-time tail in the current autocorrelations.

Writing the total current in terms of the long-wavelength density
fluctuations in reciprocal space, we have
\begin{equation}
{\bf J}_{\alpha} =
\frac{E_{\alpha\beta\gamma}-E_{\alpha\gamma\beta}}{2i}\int {\bf
q}n_{\beta}({\bf q})n_{\gamma}(-{\bf q})d{\bf
q}+...+\eta_{\alpha}~,
\end{equation}
where $n_\beta({\bf q})$ and $n_\gamma({\bf q})$ are the spatial
Fourier transforms of $n_\beta({\bf r})$and $n_\gamma({\bf r})$.
The low momentum fluctuations in $n_\beta({\bf q})$ and
$n_\gamma({\bf q})$ decay diffusively.  The eigenmodes of that
decay are the linear combinations of the densities that
diagonalize the diffusivity matrix.  What we are doing here is
considering the correction due to nonlinearity to the linear
diffusive ``fixed point'' theory.  These nonlinearities are
irrelevant in a renormalization group approach, but give singular
corrections to the leading diffusive behavior.  The long-time tail
we discuss here is one of those corrections.  At this diffusive
fixed point, in the basis of densities that diagonalizes the
diffusivity matrix, the autocorrelations of the long-wavelength
density fluctuations decay as
\begin{eqnarray}
\langle n_\beta({\bf q},0)n_\gamma({\bf q'},t)\rangle & \sim &
e^{-D_\beta q^2t}\delta_{\beta\gamma}\delta({\bf q+q'})
\label{diffu}
\end{eqnarray}
for small $q$ and large $t$, where $D_\beta$ is the diffusivity
for eigenmode $\beta$.  Thus at long times, the autocorrelation of
the total current behaves as
\begin{eqnarray}
\langle J(0)J(t)\rangle & \sim & \int d{\bf q} q^2 exp(-Dq^2 t) + \ldots  \nonumber \\
 & &\sim 1/t^{(d+1)/2}
\label{power}
\end{eqnarray}
in $d$ dimensions, where $D=D_\beta + D_\gamma$ is the sum of the
two smallest diffusivities.  Fourier transforming this to get the
conductivity, we see that for one dimension and small $\omega$,
$\sigma(\omega) \sim a - b\sqrt{|\omega|}+...$; our numerical
results (Fig.3) are quite consistent with this.  In our
finite-sized systems, the integrals on $q$ are replaced by sums,
which has the effect of rounding out this singularity at a
frequency of order $D/L^2$.  This rounding is apparent in Fig. 3,
and it can be seen to be reduced with increasing $L$.  Note that
the characteristic frequency of this diffusive finite-size effect
vanishes only as $L^{-2}$ with increasing size, in contrast to the
scale for the level repulsion, which vanishes exponentially in
$L$.

This explanation of the long-time tail in the autocorrelation of
the total current makes certain assumptions and predictions about
the dynamics and correlations of the long wavelength modes, which
we have checked numerically within our model.  First, it assumes
that the dynamics of the long-wavelength fluctuations in the
particle density $n$ and the energy density $e$ are diffusive.  We
have checked this by calculating their dynamic correlations:
$\langle|n(q,\omega)|^2\rangle$, $\langle|e(q,\omega)|^2\rangle$,
and $\langle n(q,\omega)e(-q,-\omega)\rangle$ for the chain of
length $L=18$.  At small $q$ and $\omega$ these fit quite well to
the expected diffusive behavior.  The two diffusivities differ by
approximately a factor of two.  The faster eigenmode of the
diffusion has a larger component of particle density than energy
density, and {\it vice versa} for the slower mode.

Another assumption is that the long-wavelength modes interact
nonlinearly to contribute to the total (zero momentum) current, as
given by Eq. 10 above.  For our finite-size 1d system with
discrete momenta and just the two densities, $n$ and $e$, this may
be rewritten for the particle current as
\begin{equation}
J = C\sum_{q>0} i(\sin q)(n(q)e(-q)-n(-q)e(q))+...+\eta~,
\end{equation}
where $C$ is the nonlinear coupling.  Thus we have measured $C$ to
check that it is indeed nonzero.  To do this, we define the
nonlinear combination of the densities at momentum $q$:
\begin{equation}
N(q) = i(\sin q)(n(q)e(-q)-n(-q)e(q))~,
\end{equation}
and measure its correlations with itself $\langle
N(q)N(q')\rangle$ and with the current. We find that the
correlations between $N$'s at different $q$'s are quite small
compared to the variances $\langle N^2(q)\rangle$.  To get
estimates of the coupling $C$, we thus note that
\begin{equation}
\langle N(q)J\rangle\cong C\langle N^2(q)\rangle~;
\end{equation}
this gives us an estimate of $C$ for each $q$, and we find that
these estimates agree well with one another at small $q$ and are
indeed nonzero, as expected.

The specific form of the long-time tails that we derive above are
for the case of relaxational dynamics and short-range
interactions, where the conserved densities relax diffusively.
Of course, charge carriers in real materials interact via the long-range Coulomb interaction, and the long-wavelength mode corresponding to the charge density, the plasmon, does not show diffusive relaxation; it is likely overdamped at high $T$, relaxing at a nonzero rate.
Thus conserved energy and charge are not
alone sufficient to produce the long-time tails discussed above
once Coulomb interactions (or other sufficiently long-range
interactions) are included.  What is needed is at least two
conserved densities that relax diffusively.  One possibility is
ionic conductors, where there are multiple conserved charge
carriers that can combine to produce another neutral,
diffusively-relaxing hydrodynamic mode in addition to the energy.

\section{\label{sec:conclusions}Conclusions}
We have studied the frequency-dependent conductivity of a
non-integrable fermionic chain with short range interactions. Our
studies show that such a system can be regarded as quantum chaotic
in the sense of random matrix theory, as can the current matrix
elements that go into the Kubo formula for the conductivity. We
have confirmed that charge transport in this system is not
ballistic but diffusive with non-linearities. These
non-linearities give rise to a non-analytic behavior for the
optical conductivity at low frequencies which we understand
through hydrodynamic arguments. We observe that the system shows
finite-size effects at low frequency arising from Wigner-Dyson
level repulsion and from the low momentum fluctuations of the
conserved quantities in the Hamiltonian.

The methods used to obtain these results are readily generalizable
to further explorations of the finite temperature behavior.  The
list of open questions includes other dynamic response functions
and transport coefficients (e.g. Peltier and heat conductivities),
an extension to two dimensions, a systematic investigation of the
high temperature series for transport, whose first term for the
conductivity has been computed here, and an extension to long
range interactions. We shall report on some of these in the near
future.

\begin{acknowledgments}
We are grateful to A. Lamacraft, J. Lebowitz, S. Kivelson and S. Sondhi for illuminating discussions.
The authors would also like to
thank the NSF for support through MRSEC grant DMR-0213706.
\end{acknowledgments}


\end{document}